\documentclass{INTERSPEECH2023}

\usepackage{pbox}
\usepackage{tikz}
\usepackage{pifont}
\usepackage{enumitem}
\usepackage{multirow}
\usepackage{mathtools}
\usepackage{caption,subcaption}

\newcommand{\cmark}{\ding{51}}%
\newcommand{\xmark}{\ding{55}}%

\interspeechcameraready

\title{TrustSER: On the Trustworthiness of Fine-tuning Pre-trained Speech Embeddings For Speech Emotion Recognition}
\name{Tiantian Feng, Rajat Hebbar, Shrikanth Narayanan}
\address{
  $^1$Signal Analysis and Interpretation Laboratory, Univeristy of Southern California, USA
}
\email{tiantiaf@usc.edu}

\begin{document}

\maketitle
 
\begin{abstract}

Recent studies have explored the use of pre-trained embeddings for speech emotion recognition (SER), achieving comparable performance to conventional methods that rely on low-level knowledge-inspired acoustic features. These embeddings are often generated from models trained on large-scale speech datasets using self-supervised or weakly-supervised learning objectives. Despite the significant advancements made in SER through the use of pre-trained embeddings, there is a limited understanding of the trustworthiness of these methods, including privacy breaches, unfair performance, vulnerability to adversarial attacks, and computational cost, all of which may hinder the real-world deployment of these systems. In response, we introduce TrustSER, a general framework designed to evaluate the trustworthiness of SER systems using deep learning methods, with a focus on privacy, safety, fairness, and sustainability, offering unique insights into future research in the field of SER. Our code is publicly available under: https://github.com/usc-sail/trust-ser.


\end{abstract}

\section{Introduction}
\label{section:intro}

Speech emotion recognition (SER) systems have been increasingly utilized in a wide range of ML ecosystems, including virtual assistants \cite{lee2020study}, medical diagnosis \cite{ramakrishnan2013speech, Bone2017SignalProcessingandMachine}, and education \cite{li2007speech}. Conventionally, SER systems have relied on low-level acoustic features, such as speech prosody and spectral information \cite{eyben2015geneva}. In recent years, breakthroughs in the field of deep learning have led to superior results for SER using richer time-frequency representations such as Mel-spectrograms. However, the performance of these models is often constrained by the limited size of available SER datasets. Over the past few years, large-scale pre-trained models have revolutionized speech modeling by offering effective solutions to challenging speech tasks including the SER. For example, prior works have extensively investigated models such as Wav2vec 2.0 for SER and have reported competitive results on popular SER testbeds \cite{pepino21_interspeech, chen2021exploring, wagner2022dawn}.

One of the most fundamental aspects of modern society is trust, which includes trust in AI-powered services in much the same way we trust those closest to us, such as friends, colleagues, and family members. Despite the prospect of the broad deployment of large-scale pre-trained models for SER applications, there has been limited research conducted to investigate the trustworthiness of SER systems utilizing these pre-trained models. Conventionally, ML practitioners evaluate the performance of SER applications based on the system performances, such as the unweighted average recall (UAR) score, while exploring the trustworthiness of these systems remains a largely unexplored domain.

\begin{figure}[t]
    \begin{center}
       \includegraphics[width=\linewidth]{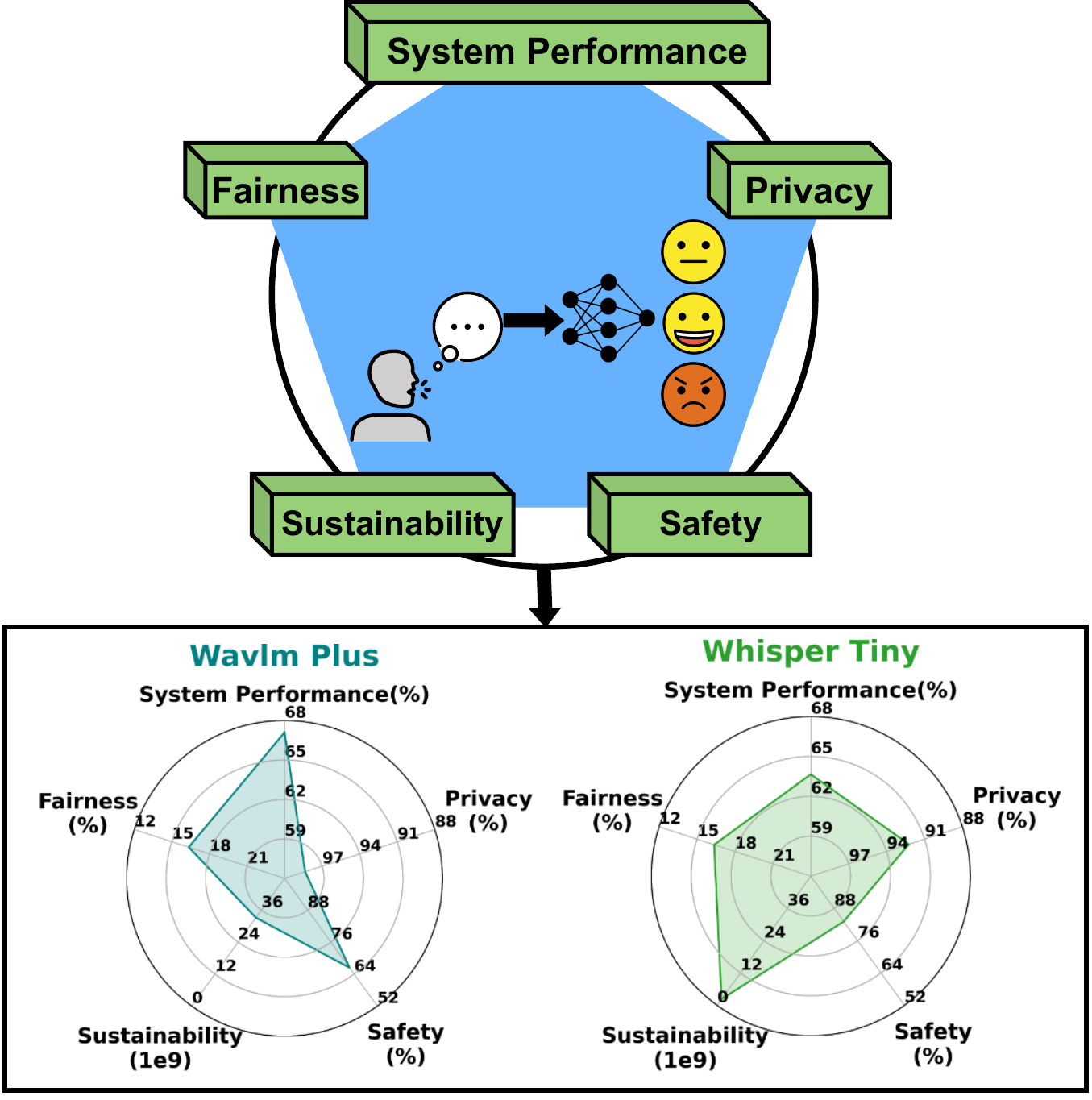}
    \end{center}
    \caption{Trustworthy profile existed in SER ecosystem: system performance, privacy, safety, fairness, and sustainability. } 
    \label{fig:trustworthy_intro}
\end{figure}

Prior literature \cite{smuha2019eu, liu2022trustworthy, feng2022review} have provided many definitions of trustworthiness, emphasizing the need for AI systems to treat people fairly, respect user privacy, act transparent, and perform consistently and efficiently. Drawing from these previous definitions, our study aims to assess trustworthiness in SER that integrates large-scale pre-trained embeddings focusing on evaluating four broad aspects of trustworthiness, including \textit{privacy}, \textit{sustainability}, \textit{safety}, and \textit{fairness}, as depicted in Figure~\ref{fig:trustworthy_intro}\footnote{The figure uses images from https://openmoji.org/ }. We highlight that the evaluation of the trustworthiness should also include the assessment of system performance.

\begin{table*}[t]
\caption{Comparison between TrustSER and existing SER studies that involve fine-tuning with pre-trained architectures.}
    \footnotesize
    \begin{tabular*}{\linewidth}{lccccccc}
        \toprule
        
        & \multirow{2}{*}{\shortstack{\textbf{System}\\\textbf{Performance}}} & 
        \multirow{2}{*}{\textbf{Fairness}} & 
        \multirow{2}{*}{\textbf{Privacy}} & 
        \multirow{2}{*}{\textbf{Safety}} & 
        \multirow{2}{*}{\textbf{Sustainability}} & 
        \multirow{2}{*}{\shortstack{\textbf{Pre-trained} \textbf{Architectures}}} &
        \multirow{2}{*}{\textbf{\#Datasets}} \\ 
        & & & & & & & \\ 

        \midrule 
        \textbf{Chen et al.} \cite{chen2021exploring} & \cmark & \xmark & \xmark & \xmark & \xmark & Wav2vec 2.0 Base & 2 \\

        \textbf{Pepino et al.} \cite{pepino21_interspeech} & \cmark & \xmark & \xmark & \xmark & \xmark & Wav2vec 2.0 Base & 2 \\

        \multirow{1}{*}{\textbf{Wagner et al.}\cite{wagner2022dawn}} & \multirow{1}{*}{\cmark} & \multirow{1}{*}{\cmark} & \multirow{1}{*}{\xmark} & \multirow{1}{*}{\xmark} & \multirow{1}{*}{\cmark} & Wav2vec 2.0 Families, HuBERT & 3 \\
        
        \midrule 
        \multirow{2}{*}{\textbf{TrustSER (Ours)}} & 
        \multirow{2}{*}{\cmark} & \multirow{2}{*}{\cmark} & 
        \multirow{2}{*}{\cmark} & \multirow{2}{*}{\cmark} & 
        \multirow{2}{*}{\cmark} & \multirow{2}{*}{\shortstack{APC, TERA, Wav2vec 2.0 Base\\WavLM Base+, Whisper Tiny,Base,Small}} & 
        \multirow{2}{*}{\textbf{4}} \\
        
        & & & & & & \\ 
        \bottomrule
    \end{tabular*}
\label{table:comparison_trust_ser}
\end{table*}

\noindent \textbf{Overview of TrustSER:} In this work, we introduce \textbf{TrustSER}, the SER benchmark that targets the evaluation with trustworthiness. Specifically, our contributions are summarized as follows:

\begin{itemize}[leftmargin=*]
    \item To the best of our knowledge, our work represents the first attempt to benchmark the trustworthiness of SER systems, covering four key components associated with trustworthy ML: \textbf{privacy}, \textbf{fairness}, \textbf{safety}, and \textbf{sustainability}.
    
    \item TrustSER evaluates on \textbf{four popular SER testbeds} using \textbf{seven representative pre-trained backbones}: APC \cite{chung2019unsupervised}, TERA \cite{liu2021tera}, Whisper Tiny, Base, Small \cite{radford2022robust}, Wav2vec 2.0 \cite{baevski2020wav2vec}, and WavLM \cite{chen2022wavlm}. 

    
    \item TrustSER proposes the \textbf{trustworthiness profile} (as shown in Fig~\ref{fig:trustworthy_intro}), a unique perspective for guiding future research directions in the field of SER. This evaluation strategy differs from the existing literature that primarily competes for state-of-the-art system performance. A more trustoworthy model profile may exhibit less skewed shape (e.g. triangular).
    
\end{itemize}

\begin{table}[t]
\caption{Summary of pre-trained encoders used in TrustSER.}
    \footnotesize
    \begin{tabular*}{\linewidth}{lcccc}
        \toprule
        
        \multirow{2}{*}{\shortstack{\textbf{Pre-trained}\\\textbf{Architecture}}} & 
        \multirow{2}{*}{\textbf{Backbone}} & 
        \multirow{2}{*}{\shortstack{\textbf{\#Layers}}} &
        \multirow{2}{*}{\shortstack{\textbf{Hidden}\\\textbf{Size}}} & 
        \multirow{2}{*}{\shortstack{\textbf{\#Params}}}  \\ 

        & & & & \\ 
         
        \midrule
        \textbf{APC} & GRU & 3 & 512 & 4.11M \\ 
        \textbf{TERA} & Transformer & 4 & 768 & 21.33M \\
        \textbf{Whisper Tiny} & Transformer & 4 & 376 & 8.21M \\ 
        \textbf{Whisper Base} & Transformer & 8 & 512 & 20.59M \\ 
        \textbf{Whisper Small} & Transformer & 12 & 768 & 88.15M \\
        \textbf{W2V 2.0 Base} & Transformer & 12 & 768 & 95.04M \\ 
        \textbf{WavLM Base+} & Transformer & 12 & 768 & 94.70M \\ 

        \bottomrule
    \end{tabular*}
\label{table:pretrained_models}
\end{table}

\section{Related Works}

\subsection{Pre-trained Speech Model}

Self-supervised learning (SSL) has emerged as a popular framework for speech representation learning, allowing for the training of large-scale data without labels. These approaches aim to learn general speech structures using generative~(\textbf{APC}\cite{chung2019unsupervised}, \textbf{TERA}~\cite{liu2021tera}), discriminative~(\textbf{Wav2vec 2.0}~\cite{baevski2020wav2vec,schneider2019wav2vec}) and multi-task learning objectives(\textbf{WavLM}~\cite{chen2022wavlm}). The resulting models often generate generic representations, making them promising candidates for fine-tuning the SER. \textbf{Whisper} \cite{radford2022robust}, on the other hand, is a transformer model trained using weakly supervised learning. We want to highlight that fine-tuning Whisper for SER is an area of limited exploration. The details of pre-trained models used in this work are summarized in Table~\ref{table:pretrained_models}.






\vspace{-2mm}
\subsection{Finetune Pre-trained Speech Models For SER} 
\vspace{-2mm}
As summarized in Table~\ref{table:comparison_trust_ser}, a number of SER frameworks that target fine-tuning existing pre-trained models have been developed in the past few years. For example, \cite{pepino21_interspeech} was one of the first SER studies to investigate fine-tuning with Wav2vec 2.0 embeddings. Their results demonstrate competitive performance compared to previous literature that did not employ pre-trained models. 
Meanwhile, \cite{chen2021exploring} proposed Task adaptive pretraining (TAPT) based on the Wav2vec 2.0 model, which further improves SER performance. 
More recently, \cite{wagner2022dawn} conducted a comprehensive analysis of fine-tuning Wav2vec 2.0 and HuBERT \cite{hsu2021hubert} for SER. Although this study is the most closely related to ours, it solely provides fairness and efficiency evaluation on fine-tuning experiments. TrustSER, on the other hand, includes additional assessments related to privacy and adversarial attacks.

\begin{figure}[t]
    \begin{center}
        \includegraphics[width=\linewidth]{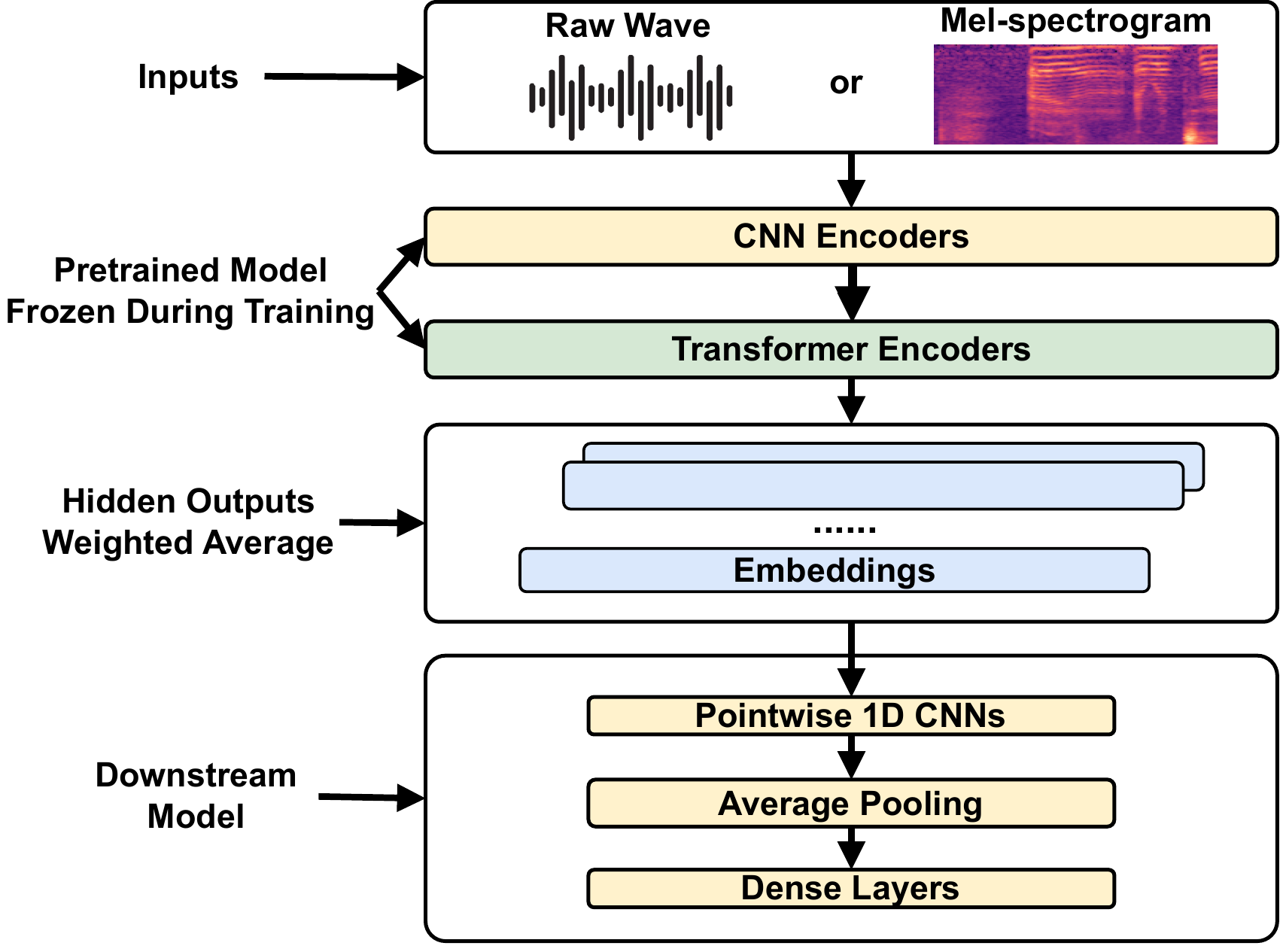}
    \end{center}
    \caption{Modeling framework used for this work. The pre-trained model shown in the diagram applies to TERA, Wav2vec 2.0, WavLM, and Whisper models. However, the pre-trained encoders would be substituted with GRU layers for APC.} 
    \label{fig:downstream_model}
\end{figure}

\section{Method}

In this section, we describe our fine-tuning model architecture and trustworthiness evaluation metrics used in TrustSER. The complete architecture is illustrated in Fig~\ref{fig:downstream_model}.

\vspace{-1mm}
\subsection{Modeling Methods}
\vspace{-1mm}

\noindent \textbf{Frozen vs. Unfrozen Pre-trained Models}: Our modeling approach draws inspiration from \cite{pepino21_interspeech}, which highlights that fine-tuning while keeping the backbone encoder frozen provides simple but competitive results compared to unfreezing the encoder. Additionally, \cite{pepino21_interspeech} demonstrate that using the hidden outputs from all encoder layers outperforms using solely the hidden output from the last layer for the downstream SER task. Consequently, we adopt a similar approach of freezing all the pre-trained models and utilizing the hidden output from all encoders for fine-tuning the downstream SER task.

\begin{figure*}[t]
    \begin{center}
        \includegraphics[width=\linewidth]{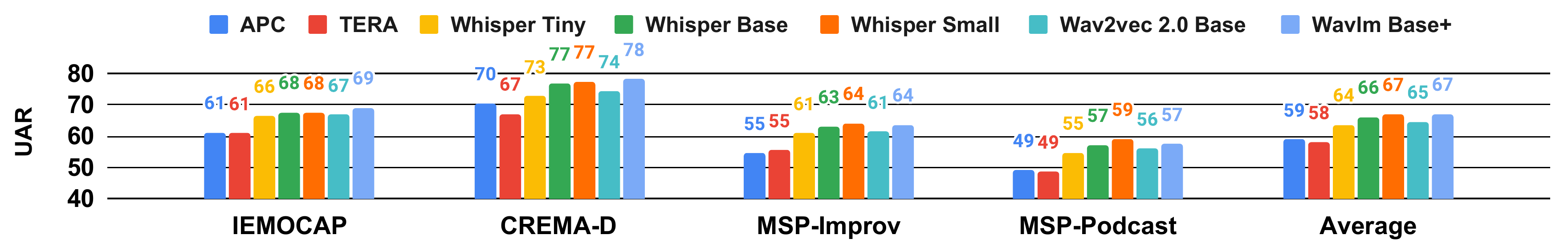}
    \end{center}
    \caption{System performance with fine-tuning different pre-trained models (frozen during training) for SER.} 
    \label{fig:system_performance}
\end{figure*}

\noindent \textbf{Downstream Model}: Our downstream model starts with a weighted averaging to combine the hidden outputs from all encoder layers, where the weights are parameterized. We then pass the resulting weighted average output to two 1D pointwise convolutional layers with a kernel size of 1 and a filter size of 128, with the ReLU activation functions in between. The outputs from the convolutional layers are then averaged over the timestamps to yield a vector of size 128. Finally, this vector is fed into two fully connected layers for the SER prediction.

\vspace{-2mm}
\subsection{Trustworthiness Evaluation}
\vspace{-1mm}
We provide a summary of the trustworthiness with regard to privacy, safety, fairness, and sustainability below:

\noindent \textbf{System Performance} serves as the fundamental metric to evaluate an SER system. Due to the imbalanced data distribution in SER datasets, the conventional system performance metric used in SER is unweighted average recall (UAR).

\noindent \textbf{Privacy:} Speech often carries sensitive information about individuals, exposing users to significant privacy risks \cite{liu2021machine}. This study focuses on the property inference attack, where adversaries attempt to infer speaker demographics that is unrelated to the SER task. To assess the privacy of pre-trained embeddings, we employ gender inference as adopted in \cite{feng2022enhancing}. More results related to speaker recognition would be located in our GitHub with ongoing releases.

\noindent \textbf{Safety:} Prior research in \cite{wagner2022dawn} has explored the robustness of fine-tuning pre-trained models for SER by applying Gaussian noises and environmental noises to the raw waveform. However, adversarial attacks have emerged as a more effective approach to destabilize an SER system \cite{latif2020deep}. The fast gradient sign method (FGSM) is a widely used method to construct adversarial examples where a perturbation is applied to a data point $x$ \cite{goodfellow2014explaining}:

\begin{equation}
    \delta = \epsilon * \mathrm{sign}(\nabla_x \mathcal{L}(x,y;\theta ))
\end{equation}

where $\mathcal{L}(\cdot)$ represents the loss function, $y$ is the label, $\theta$ is the model parameters, and $\epsilon$ constrains the amount of perturbation. TrustSER also supports other adversarial attacks like Project Gradient Descent (PGD) attacks \cite{madry2017towards}, but due to the limited space, we focus on evaluating the robustness and safety of the SER system against adversarial attacks using FGSM.

\noindent \textbf{Fairness:} Given the growing concern over societal biases and imbalanced data attributes in ML systems, it is critical to ensure that such systems do not exacerbate existing unfairness issues in the ecosystem. To evaluate fairness, we utilize Equality of odds as suggested by the prior SER work \cite{gorrostieta2019gender}, a metric that measures whether the model predictions are unbiased with respect to a protected variable or not. In this work, we choose gender as the protected variable. TrustSER also supports other fairness metrics like statistical parity on our GitHub repo.

\vspace{0.6mm}

\noindent \textbf{Sustainability:} In addition to fairness, privacy, and safety, efficiency is an essential aspect to consider when deploying SER systems in real-world applications. While prior works \cite{chen2021exploring, wagner2022dawn} have investigated the data efficiency of fine-tuning frameworks for SER, this study specifically focuses on inference computation cost that is closely related to the feasibility of deploying the system. More precisely, our goal is to quantify the computational resources during the inference stage using floating-point operations per second (FLOPs).

\noindent \textbf{Trustworthiness Profile:} As shown in Fig~\ref{fig:trustworthy_intro}, the trustworthy profile integrates evaluation in all trustworthy aspects and summarize them into one model profile. The profile provides direct information in choosing modeling approach that best suits for the deployment. For example, in a mobile deployment environment, ML practitioners may choose the model with more focus on sustainability rather than best system performance.

\begin{table}[t]
\caption{Summary of dataset statistics used in this work.}
    \footnotesize
    \begin{tabular*}{\linewidth}{lccccc}
        \toprule
        
        \multirow{1}{*}{\shortstack{\textbf{Datasets}}} & 
        \multirow{1}{*}{\textbf{Neutral}} & 
        \multirow{1}{*}{\shortstack{\textbf{Happy}}} &
        \multirow{1}{*}{\shortstack{\textbf{Sad}}} & 
        \multirow{1}{*}{\shortstack{\textbf{Angry}}} & 
        \multirow{1}{*}{\shortstack{\textbf{Total}}}  \\ 
         
        \midrule
        \textbf{IEMOCAP} & 1,708 & 1,636 & 1,084 & 1,103 & 5,531 \\ 
        \textbf{CREMA-D} & 1,972 & 1,219 & 588 & 1,019 & 4,798 \\ 
        \textbf{MSP-Improv} & 3,477 & 2,644 & 885 & 792 & 7,798 \\ 
        \textbf{MSP-Podcast} & 20,986 & 12,060 & 2,166 & 2,712 & 37,924 \\
        \midrule
        \textbf{Total} & 28,143 & 17,559 & 4,723 & 5,716 & 56,051 \\

        \bottomrule
    \end{tabular*}
\label{table:datasets}
\end{table}

\section{Datasets}

Table~\ref{table:datasets} displays data statistics for the four datasets included in TrustSER. Due to the existence of imbalanced label distribution within the dataset, we decided to keep the four most frequently presented emotions for all the datasets, as recommended in \cite{feng2022enhancing,feng2021privacy,chen2021exploring,pepino21_interspeech}. Below is a brief overview of each included dataset:

\noindent \textbf{IEMOCAP} \cite{busso2008iemocap} contains multi-modal (motion, audio, and video) recordings of acted human interactions from ten subjects, evenly distributed between males and females.

\noindent \textbf{CREMA-D} dataset consists of audio-visual clips that were recorded using 91 actors \cite{cao2014crema}, who were directed to express six specific emotions while uttering a set of 12 sentences. 

\noindent \textbf{MSP-Improv}~\cite{busso2016msp} corpus is developed with the target of investigating naturalistic emotions that were elicited from improvised situations. The corpus is comprised of both audio and visual data collected from 12 individuals, with an equal number of subjects from both male and female participants.

\begin{figure*}[t] {
    \centering
    
    \vspace{-3mm}
    \begin{tikzpicture}
        
        \node[draw=none,fill=none] at (0,4){\includegraphics[width=0.245\linewidth]{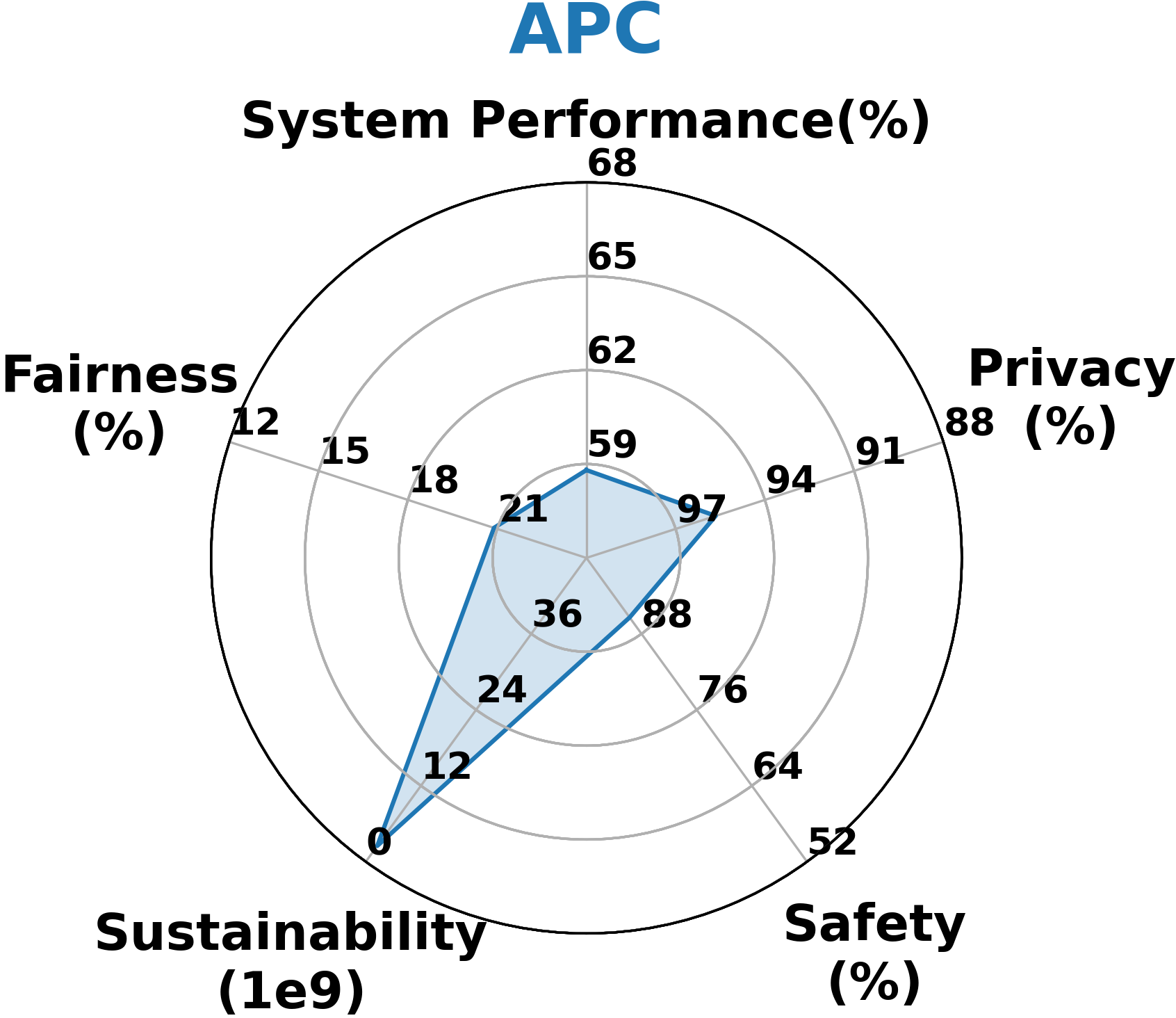}};

        \node[draw=none,fill=none] at (0.25\linewidth,4){\includegraphics[width=0.245\linewidth]{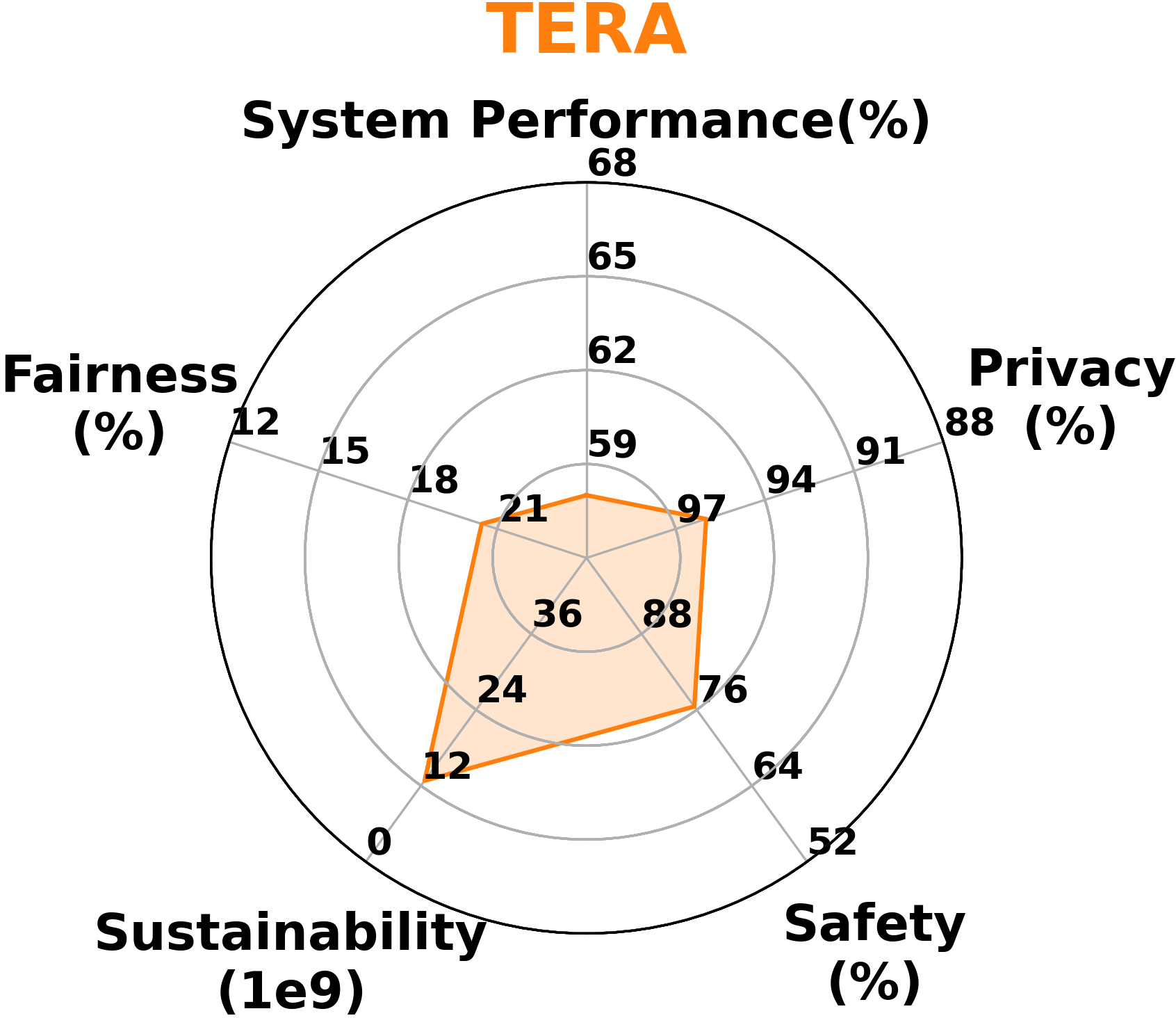}};
        
        \node[draw=none,fill=none] at (0.5\linewidth,4){\includegraphics[width=0.245\linewidth]{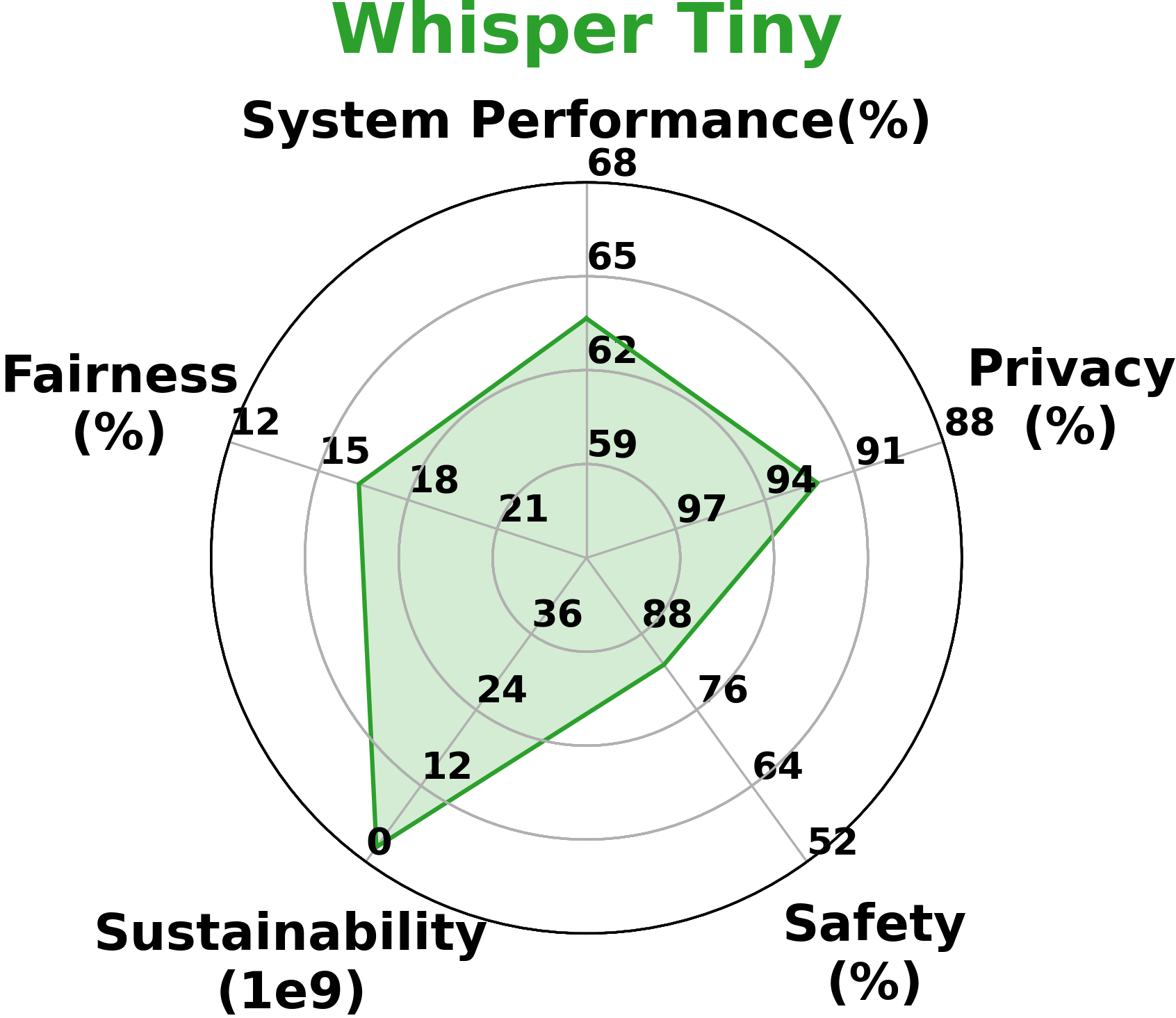}};

        \node[draw=none,fill=none] at (0.75\linewidth,4){\includegraphics[width=0.245\linewidth]{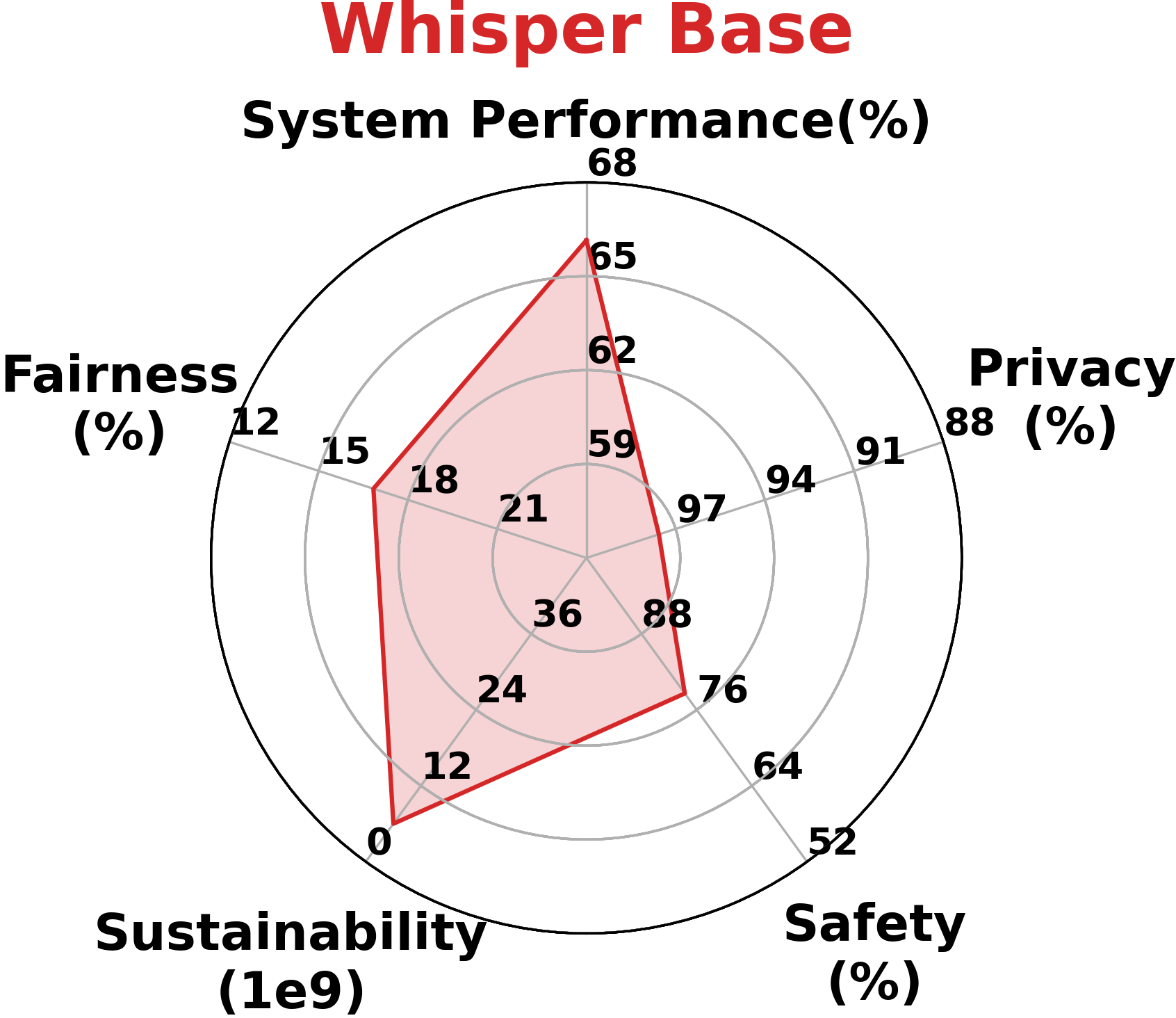}};

        \node[draw=none,fill=none] at (0,0){\includegraphics[width=0.245\linewidth]{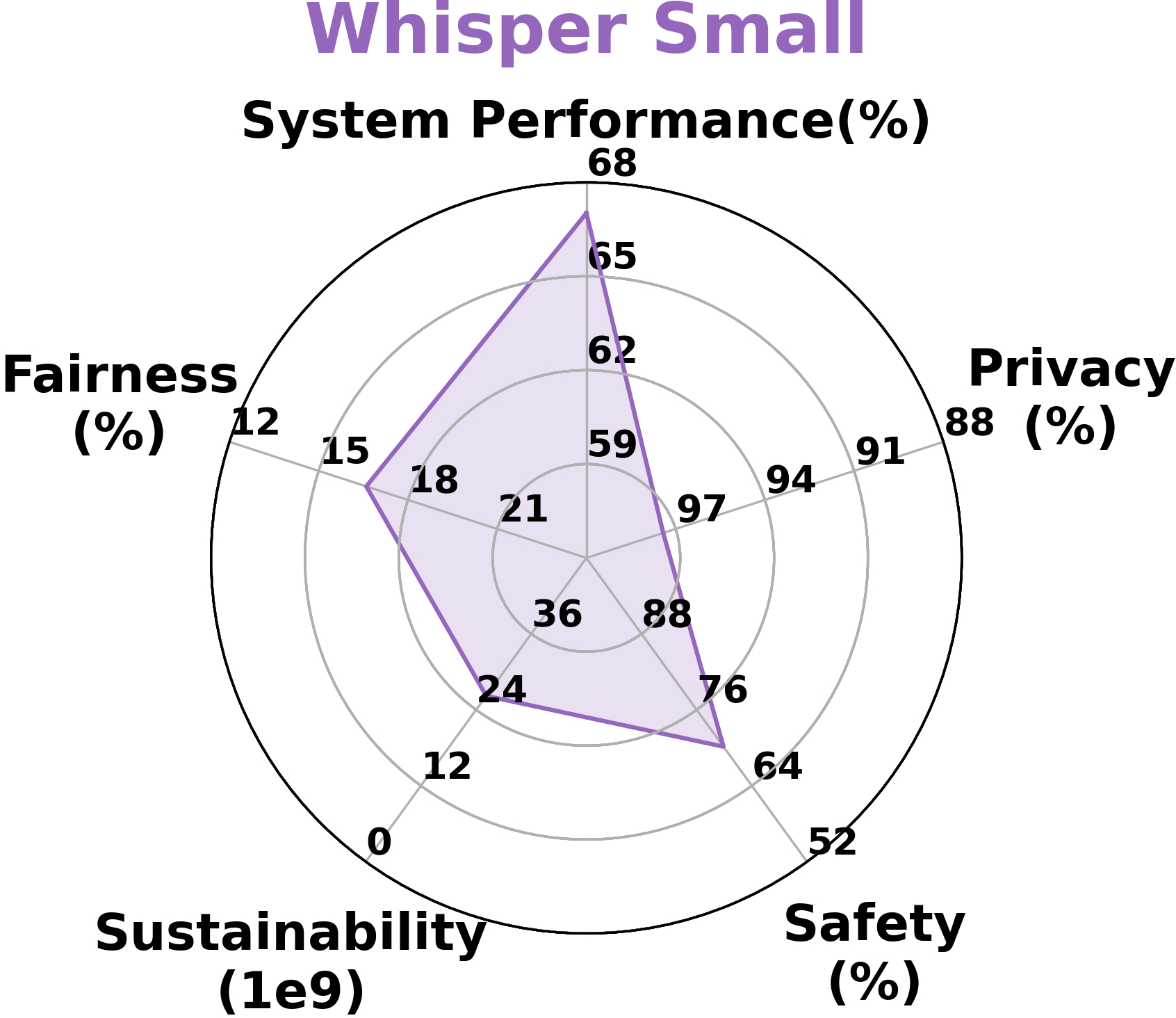}};
        
        \node[draw=none,fill=none] at (0.25\linewidth,0){\includegraphics[width=0.245\linewidth]{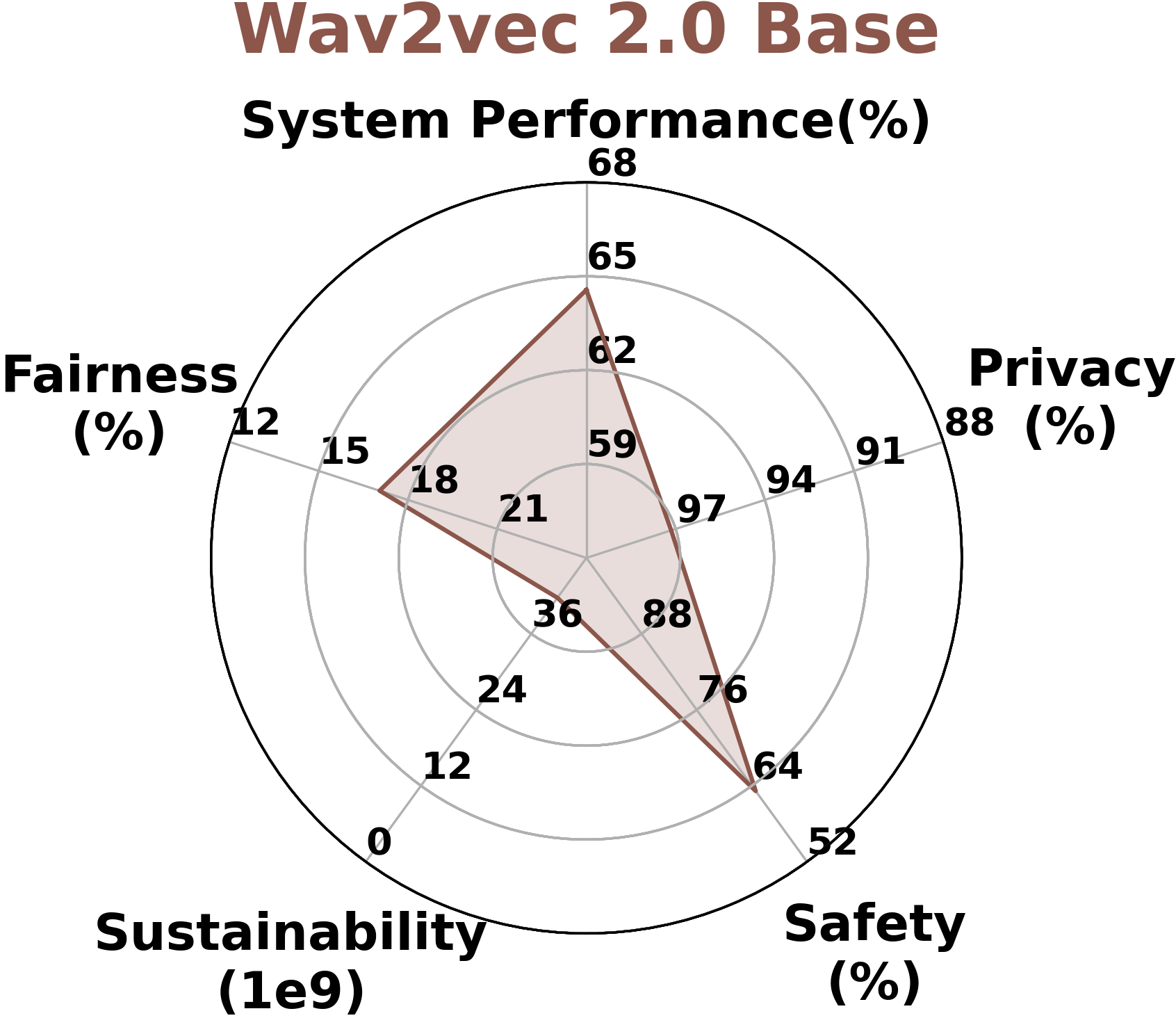}};

        \node[draw=none,fill=none] at (0.5\linewidth,0){\includegraphics[width=0.245\linewidth]{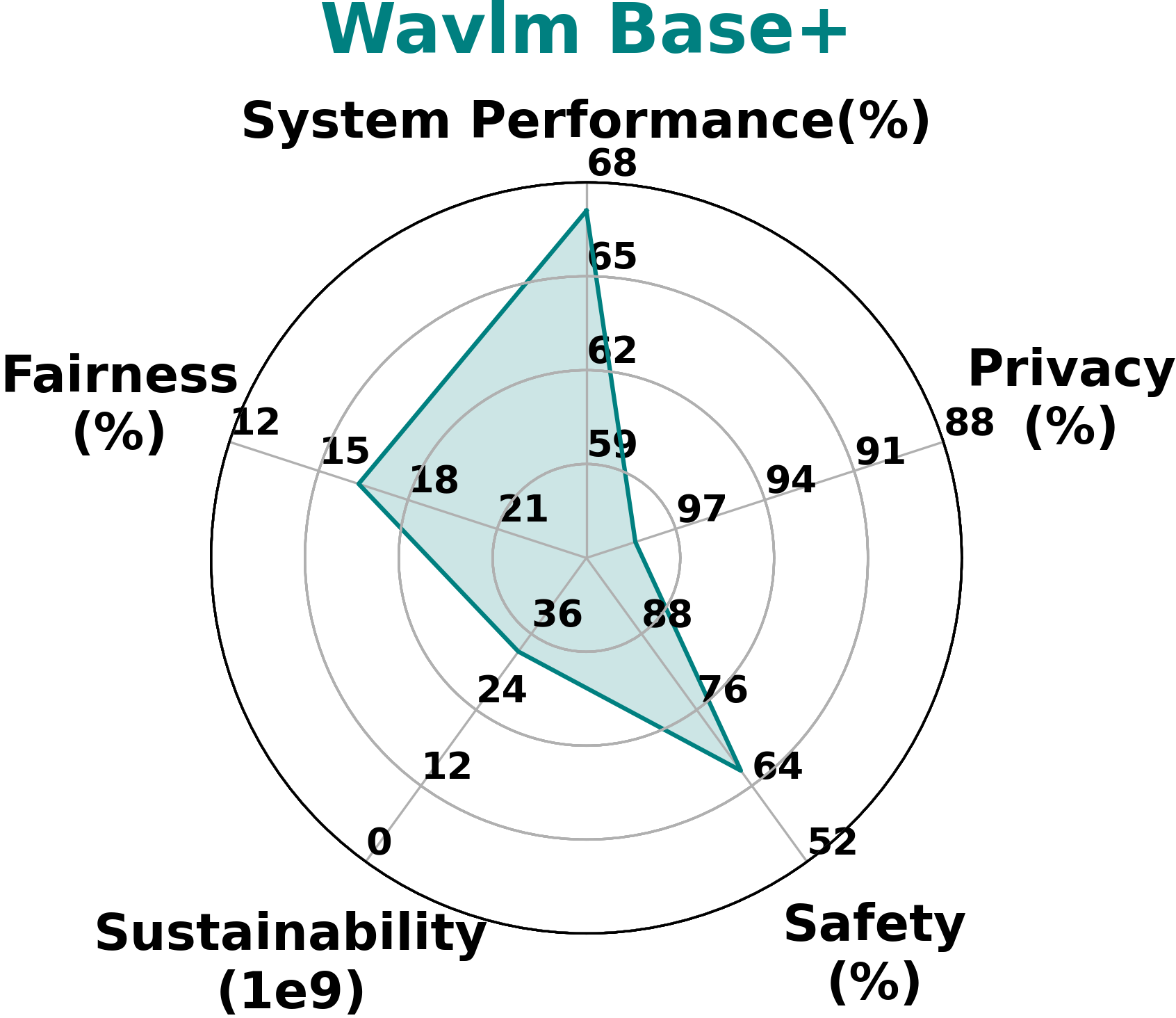}};

        \node[draw=none,fill=none] at (0.75\linewidth,0){\includegraphics[width=0.245\linewidth]{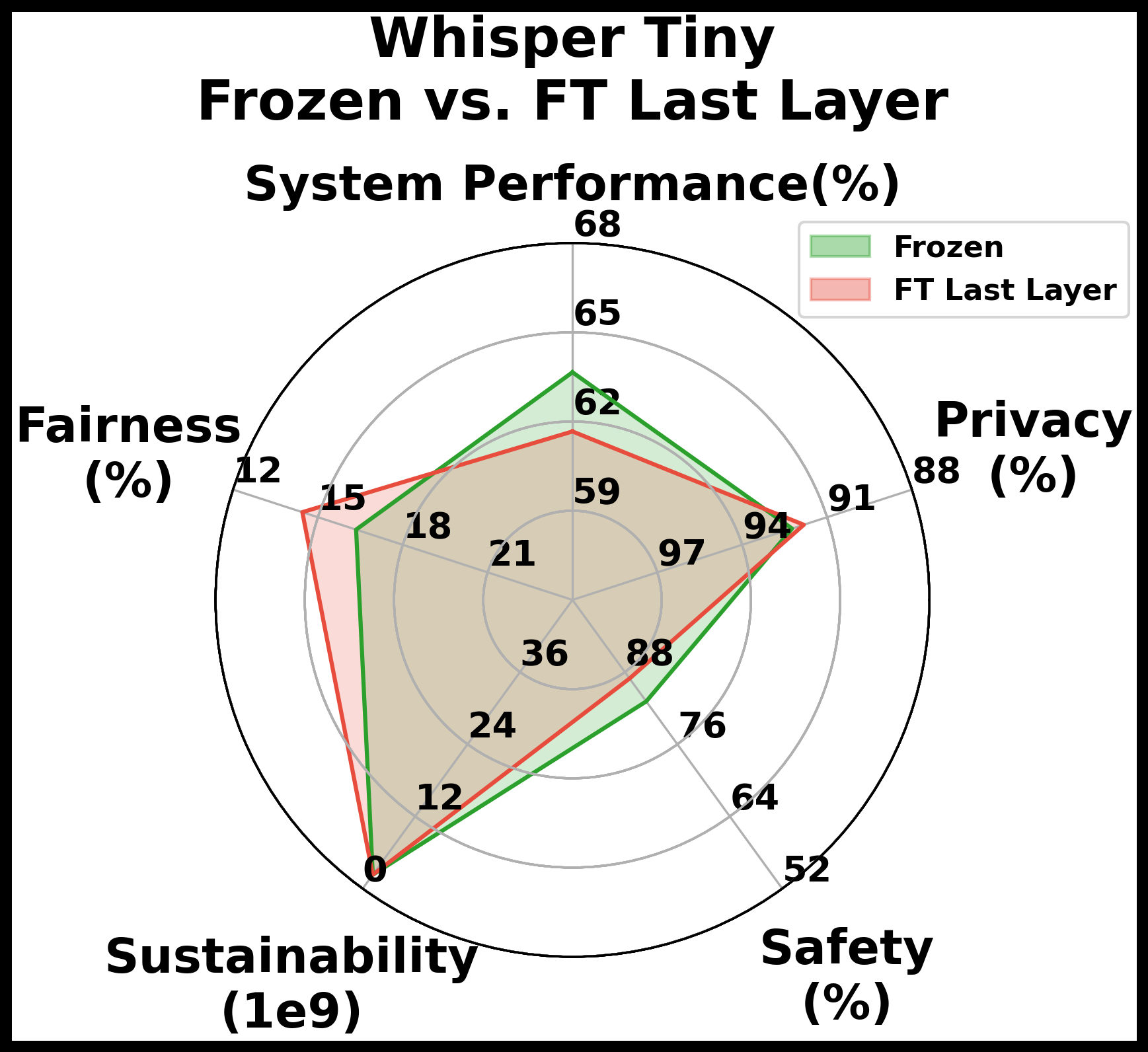}};
        
    \end{tikzpicture}
    \caption{Trustworthy profile for SER using pre-trained models. The scores are backward for privacy, fairness, sustainability, and safety, as lower scores indicate better metrics. We also show a case study comparing: fine-tuning last layer vs. freezing Whisper Tiny.}
    \label{fig:trust_profile}
    
} \end{figure*}

\noindent \textbf{MSP-Podcast}~\cite{lotfian2017building} is collected from podcast recordings, where there are 610 speakers in the training split, 30 in the development, and 50 speakers in the test split. 

\section{Results}

\subsection{Experiment Details}

\noindent \textbf{Data Split}: We apply 5-fold and 6-fold evaluation on IEMOCAP and MSP-Improv datasets, where each session is regarded as a unique test fold. During each training fold, one session of data is used for validation while the rest are for training. Moreover, we perform 5-fold experiments on the CREMA-D dataset, where 20\% of speakers are used as testing data in each fold. On the other hand, we used the standard splits for training, validation, and testing from the MSP-Podcast dataset.

\noindent \textbf{SER Training}: We set the batch size as 64 in all fine-tuning experiments. Specifically, we set the learning rate as 0.0005 and the maximum training epoch as 30. We set the maximum audio duration as 6 seconds. All experiments are implemented using PyTorch. The experiments are conducted on a computing server with two RTX 5000 GPUs. We use the checkpoints for APC and TERA from Superb \cite{yang21c_interspeech}, and the remaining models use the checkpoints from HuggingFace \cite{wolf-etal-2020-transformers}.

\vspace{-1mm}
\subsection{System Performance}
\vspace{-1mm}

Figure~\ref{fig:system_performance} displays the System performance with fine-tuning different pre-trained embeddings for SER. From the plot, we can find that the APC and TERA models provide subpar performance on most SER datasets. In contrast, the large-scale transformer model WavLM achieves the best performance on average on multiple SER datasets. Meanwhile, Whisper-Small, which has a similar number of parameters, achieves comparable system performance to WavLM. However, Wav2vec 2.0 does not provide competitive results compared to WavLM and Whisper-Small.  Surprisingly, our findings indicate that the parameter-light Whisper-tiny model produces comparable results to Wav2vec 2.0 despite having only around~8M parameters. In summary, we find that larger transformers generally provide better performances compared to their smaller counterparts, although the difference in performance is not substantial.


\subsection{Trustworthiness Evaluation}

We summarize the evaluation of the trustworthiness between different pre-trained architectures in Table~\ref{table:trustworthiness}. The scores reported are average values across all the datasets.

\vspace{1mm}
\noindent \textbf{Trustworthiness Profile:} We present the trustworthy profiles of used pre-trained embeddings in Fig~\ref{fig:trust_profile}. Results suggest that larger transformer models tend to have profiles that favor system performance but come with higher computational costs and privacy risks. Meanwhile, smaller models such as Whisper-Tiny achieve comparable performance with fewer computations and lower privacy concerns while posing safety risks for real-world applications in the presence of adversarial attacks. The details analysis of individual trustworthy elements is presented in the following subsections.

\vspace{1mm}
\noindent \textbf{Privacy}: We use the same fine-tuning approach for gender prediction and we report the accuracy for evaluating the privacy of pre-trained model architecture. We use a learning rate of 0.0005, while the maximum training epoch is set to 10, as we find fast convergence in gender prediction. From the table, we can observe that most pre-trained embeddings are indicative of gender, and models with fewer parameters show slightly lower performance for gender classification. This suggests the potential privacy risks posed by pre-trained embeddings.

\vspace{1mm}
\noindent \textbf{Safety}: We constrain the $\epsilon$ to apply an SNR=$45$dB in creating adversarial speech examples (more results regarding adversarial noise at different SNRs and comparisons between adversarial noise and Gaussian noise can be found on our GitHub). The adversarial attack is applied to all the test samples with correct predictions, and we use the \textit{attack success rate (ASR)} to evaluate the adversarial attacks. Table~\ref{table:trustworthiness} exhibits that fine-tuned SER systems are vulnerable to adversarial attacks, with the lowest attack success rate at 53.9\%. Moreover, we can observe that models with more parameters are more robust against adversarial attacks, complying with the findings reported in \cite{bhojanapalli2021understanding}.

\begin{table}[t]
\caption{Trustworthiness evaluation results: lower scores correspond to better metrics. ASR is the attack success rate.}
    \footnotesize
    \begin{tabular*}{\linewidth}{lcccc}
        \toprule
        
        \multirow{2}{*}{\shortstack{\textbf{Pre-trained}\\\textbf{Model}}} & 
        \multirow{2}{*}{\shortstack{\textbf{Gender}\\\textbf{Pred. (\%)}}} & 
        \multirow{2}{*}{\shortstack{\textbf{ASR }\\\textbf{(\%)}}} &
        \multirow{2}{*}{\shortstack{\textbf{Equality }\\\textbf{of Odds (\%)}}} & 
        \multirow{2}{*}{\shortstack{\textbf{FLOPs}\\($\mathbf{10^9}$)}} \\ 

        & & & & \\ 
         
        \midrule
        \textbf{APC} & 95.6 & 88.2 & 20.9 & 2.5 \\ 
        \textbf{TERA} & 95.7 & 70.7 & 20.5 & 12.8 \\ 
        \textbf{Whisper Tiny} & \textbf{92.2} & 78.9 & 16.6 & \textbf{2.3} \\ 
        \textbf{Whisper Base} & 97.6 & 73.2 & 17.0 & 6.0 \\ 
        \textbf{Whisper Small} & 97.4 & 61.0 & 16.6 & 26.2 \\
        \textbf{W2V 2.0 Base} & 97.2 & \textbf{53.9} & \textbf{16.4} & 41.7 \\ 
        \textbf{WavLM Base+} & 98.4 & 57.9 & 16.8 & 33.2 \\ 

        \bottomrule
    \end{tabular*}
\label{table:trustworthiness}
\end{table}

\vspace{1mm}
\noindent \textbf{Fairness}: We compute the average Equality of odds from all datasets. A higher score corresponds to a more biased model performance. Based on the results presented in Table~\ref{table:trustworthiness}, it can be discovered that transformer-based models demonstrate better fairness scores compared to the model based on GRUs. Notably, the fairness scores for the Wav2vec 2.0, WavLM, and Whisper model families appear to be similar. Furthermore, our results indicate that models with a smaller number of parameters do not necessarily exhibit significantly better fairness metrics than those with a higher number of parameters.

\vspace{1mm}
\noindent \textbf{Sustainability}: We evaluate sustainability using the FLOPs during the inference stage, where we set the audio length to be 6 seconds. The FLOPs data demonstrate a significantly higher computational cost for Wav2vec 2.0 base, WavLM base+, and Whisper Small than the remaining models. Although APC has fewer parameters than Whisper-Tiny, the FLOPs show similar computation costs between these two models.

\vspace{1mm}
\noindent \textbf{Case Study on Fine-tuning Whisper Tiny:} We perform a case study with fine-tuning the last layer of Whisper-Tiny for SER, as shown in Fig~\ref{fig:trust_profile}. Our results align with previous findings \cite{pepino21_interspeech} that fine-tuning pre-trained models can reduce performance. However, the trustworthy profile suggests that fine-tuning can increase fairness, at the cost of reduced robustness to adversarial attacks. This may be because pre-trained transformers have a more generalized representation, which is lost when fine-tuned.

\begin{table}[t]
\caption{Potential design priorities for SER systems in different computing scenarios.}
    \footnotesize
    \begin{tabular*}{\linewidth}{lccc}
        \toprule
        
        \multirow{2}{*}{\shortstack{\textbf{}\\\textbf{}}} & 
        \multirow{2}{*}{\shortstack{\textbf{Edge}\\\textbf{Device}}} & 
        \multirow{2}{*}{\shortstack{\textbf{Cloud}\\\textbf{System}}} & 
        \multirow{2}{*}{\shortstack{\textbf{Critical}\\\textbf{Facilities}}}  \\ 

        & & & \\ 
        
        \midrule
        \textbf{System Performance} & Med & \textbf{High} & \textbf{High} \\ 
        \textbf{Safety} & Low & Med & \textbf{High} \\
        \textbf{Privacy} & Low & \textbf{High} & Low \\ 
        \textbf{Sustainability} & \textbf{High} & Low & Low \\ 
        \textbf{Fairness} & Med & \textbf{High} & Med \\
        \midrule
        
        \multirow{2}{*}{\textbf{Model Candidate}} & 
        \multirow{2}{*}{\shortstack{\textbf{Whisper}\\\textbf{Tiny}}} & 
        \multirow{2}{*}{\shortstack{\textbf{Whisper}\\\textbf{Small}}} & 
        \multirow{2}{*}{\textbf{WavLM Base+}}  \\ 
        & & & \\ 

        \bottomrule
    \end{tabular*}
\label{table:practical_usage}
\end{table}

\section{Practical Uses of Trust-SER}

Apart from computing the trust profile for the SER model, the major goal of Trust-SER is to provide effective guidance for ML practitioners to select the most suitable pre-trained architecture according to their requirements. For example, we summarize the potential design priorities in three usage scenarios shown in \ref{table:practical_usage}: edge computing, cloud systems, and critical facilities. However, it is worth noting that these design considerations might still vary in specific applications under these scenarios.

\vspace{1mm}
\noindent \textbf{Edge computing}: To begin with, it is crucial to achieve a satisfactory balance between system performance and efficiency in edge computing applications due to the contained computing resources available. Therefore, researchers typically sacrifice the system performance for computational efficiency in edge computing applications. Conversely, as the edge device typically performs the complete inference locally, concerns regarding privacy and safety are relatively lower, as the data never leaves the device. Based on these considerations, ML researchers can opt for a model like \texttt{Whisper-tiny} in this use case.

\vspace{1mm}
\noindent \textbf{Cloud computing}: Unlike edge devices, cloud computing facility has fewer limitations in computation resources. Hence, it is more critical to ensure system performance and metrics like fairness over sustainability. However, as the user data is often transferred to the cloud for processing, protecting user privacy becomes equally important, such as minimizing the inference of private attributes through pre-trained embeddings. Consequently, the optimal model selection could be \texttt{Whisper-Small} based on the results presented in this paper.

\vspace{1mm}
\noindent \textbf{Critical Facility}: Researchers frequently need to prioritize safety and system performance for critical facilities, as applications running on these facilities require robust and precise predictions. In contrast, privacy and sustainability are given comparatively lower priority in this scenario. As a result, researchers can select \texttt{WavLM Base+} as the pre-trained model for their SER applications.

\vspace{1mm}
\noindent \textbf{Summary:} These examples demonstrate the effectiveness of the trustworthiness profile for ML practitioners in selecting an appropriate model that meets specific requirements.

\section{Conclusion}


In recent years, deep learning have significantly improved SER performance, while also raising concerns about trustworthiness in sustainability, fairness, privacy, and safety. To address these concerns, we propose an SER benchmark called TrustSER, which evaluates the trustworthiness of SER systems using pre-trained models. Our benchmark shows that improved system performance comes at a significant computational cost, potentially hindering the broader deployment. However, larger models can enhance safety against adversarial attacks. Future research could investigate the impact of improving privacy, fairness, and safety on other aspects of trustworthiness. Other directions would be investigating trustworthy profiles under Federated Learning scenarios \cite{zhang2023fedaudio}.

\section{Acknowledgement}
This work was supported by USC Amazon Center for Secure and Trusted Machine Learning.

\bibliographystyle{IEEEtran}
\bibliography{mybib}

\end{document}